\documentclass[preprints,article,accept,moreauthors,LaTex,pdftex]{Definitions/mdpi} 

\firstpage{1} 
\pubvolume{xx}
\issuenum{xx}
\articlenumber{xx}
\pubyear{2020}
\copyrightyear{2020}
\history{Received: date; Accepted: date; Published: date}

\usepackage{gensymb}
\usepackage{epstopdf}

\Title{Electro-optical ion trap for experiments with atom-ion quantum hybrid systems}

\Author{Elia Perego $^{1, *}$, Lucia Duca $^{1}$ and Carlo Sias $^{1, 2, 3}$}

\AuthorNames{Firstname Lastname, Firstname Lastname and Firstname Lastname}

\address{%
$^{1}$ \quad Istituto Nazionale di Ricerca Metrologica, I-10135 Torino, Italy\\
$^{2}$ \quad European Laboratory for Nonlinear Spectroscopy (LENS), I-50019 Sesto Fiorentino, Italy\\
$^{3}$ \quad Istituto Nazionale di Ottica-CNR, I-50019 Sesto Fiorentino, Italy}

\corres{Correspondence: e.perego@inrim.it}

\abstract{%
In the development of atomic, molecular and optical (AMO) physics, atom-ion hybrid systems are characterized by the presence of a new tool in the experimental AMO toolbox: atom-ion interactions. 
One of the main limitations in state-of-the-art atom-ion experiments is represented by the micromotion component of the ions' dynamics in a Paul trap, as the presence of micromotion in atom-ion collisions results in a heating mechanism that prevents atom-ion mixtures from undergoing a coherent evolution.
Here we report the design and the simulation of a novel ion trapping setup especially conceived for the integration with an ultracold atoms experiment. %
The ion confinement is realized by using an electro-optical trap based on the combination of an optical and an electrostatic field, so that no micromotion component will be present in the ions' dynamics.  
The confining optical field is generated by a deep optical lattice created at the crossing of a bow-tie cavity, while a static electric quadrupole ensures the ions' confinement in the plane orthogonal to the optical lattice. 
The setup is also equipped with a Paul trap for cooling the ions produced by photoionization of a hot atomic beam, and the design of the two ion traps facilitates the swapping of the ions from the Paul trap to the electro-optical trap. 
}


\begin{document}

\section{Introduction}

Recently, the possibility of merging together ultracold atoms and trapped ions in the same experiment for creating a hybrid atom-ion quantum system has risen a lot interest \cite{Tomza-review}. 
In fact, this composite system inherits the properties of its constituents, i.e. the possibility of controlling  space-localized trapped ions and creating large ensambles of coherent matter with quantum gases, and makes it possible to use a new feature: atom-ion interactions. Atom-ion interactions are much longer-ranged than atom-atom interactions (if Rydberg states are not considered), and can in principle be tuned via Feshbach resonances \cite{Idziaszek-feshbach}.
Systems of ultracold atoms and trapped ions represent an extremely versatile and powerful experimental platform, in which both atoms and ions can simultaneously play the role of probe and system \cite{Sias.atom.ion}.
For instance, a single ion immersed in an ultracold bath of neutral atoms can be conceived as a localized impurity in a many-body system. 
In this scenario, the ion can be exploited as a probe for measuring local properties of the atomic gas, like its densities or correlations \cite{Kollath}. 
Analogously, the atomic cloud at low temperatures can act as an ultracold buffer gas that continuously cools the ions through elastic collisions, realizing a transparent coolant at the wavelengths used to manipulate the internal states of the ions \cite{Zipkes-nature}.
Ultracold buffer gas cooling can thus be an efficient alternative to laser cooling for preparing ion crystals in quantum technology applications.
Atom-ion hybrid systems are also suitable for studying in a controlled way chemical reactions at low temperature, since the ions increase the local density of the neutral atoms cloud, thus acting as reaction centers for chemical processes, e.g. charge-exchange and molecule formation \cite{Ratschbacher-chemical, Hall-charge-exchange}.

In most atom-ion experiments, the collisional energy is set by the driven component of the ions’ dynamics in radiofrequency (RF) traps (i.e., micromotion), since collisions with cold atoms cause a coupling of energy from the driving field to the ions secular motion, thus making the systems energetically open \cite{Harter}.
A relatively large collisional energy represents a limitation for experiments aiming at exploiting the quantum nature of atom-ion interactions, which arise mainly when atom-ion interactions can be characterized solely by the s-wave phase shift, i.e. the scattering length. 
In this so-called s-wave scattering regime, the collisions could be controlled via Feshbach resonances and atom-ion interactions could lead to a coherent evolution of the two quantum systems \cite{Sias.atom.ion}.
However, this regime has not been reached so far due to the presence of micromotion, and Feshbach resonances have never been experimentally observed in atom-ion mixtures.
A possible solution to reach the quantum regime with RF-trapped ions in an ultracold bath of atoms is to properly choose the atomic species on the basis of their mass ratio, thus limiting the energy transfer from the RF trapping field to the atom-ion system due to the presence of micromotion, and lowering the experimentally attainable temperature \cite{Cetina-vuletic}. 
Following this strategy, the group of R. Gerritsma has recently observed atom-ion collisional energies for which only s- and p-waves contribute to the collisions \cite{Gerritsma-quantum-regime}.
An alternative strategy to circumvent the detrimental effects of micromotion is to change the ion confinement technique. For instance, ions can be trapped in optical dipole traps \cite{Schneider-optical-ion-trapping, Schaetz-fort}. 
Purely optical trapping of a single Barium ion has been realized for up to a few seconds with a trapping scheme based on two lasers having different frequencies \cite{Schaetz-long-trapping}.
Moreover, optical trapping has also been achieved by employing a one-dimensional optical lattices formed by two counter-propagating beams \cite{Schaetz-paul-lattice}, and very recently small crystals of up to six Barium ions were trapped in a single-beam dipolar trap \cite{Schaetz-crystal}. 
Hence, ion optical trapping represents an encouraging solution to reach ultracold atom-ion collisions.

In this paper, we present the design of a novel ion trap made of a combination of optical and electrostatic fields for trapping two-dimensional crystals of ions. 
This electro-optical (EO) trap is based only on static fields, so no micromotion will affect the ions' dynamics in the trap. 
The trap is part of an atom-ion experiment that aims at observing collisions between Ba$^+$ ions and Li atoms in the so-far elusive s-wave regime \cite{Perego}.



\section{The electro-optical trap}

\begin{figure}[t]
\centering
\includegraphics[width=0.7\textwidth]{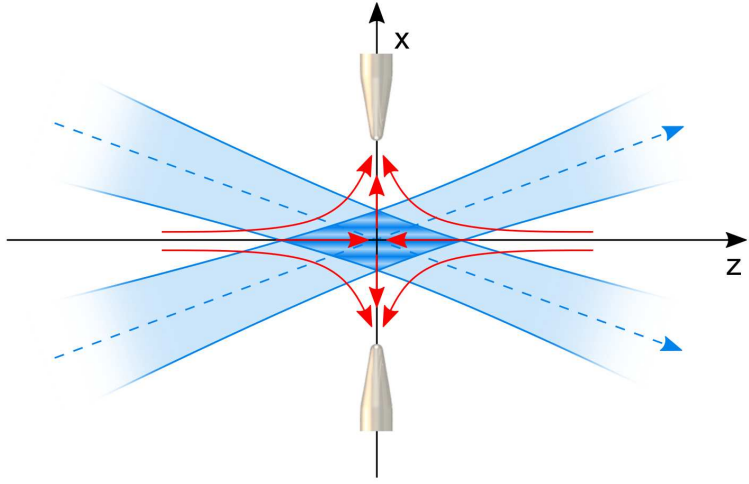}
\caption{Sketch of the electro-optical trap, in which a static electric quadrupole potential (red arrows) is used for trapping ions along two orthogonal directions (y and z axes), while the confinement along the antitrapping direction (x axis) is provided by the interference pattern of two crossed gaussian laser beams (drawn in blue). For positive ions, the electrodes generating the quadrupole potential must be negatively charged.
\label{fig:EoScheme}}
\end{figure}

The electro-optical potential is created by the superposition of a one-dimensional optical lattice and a static electric quadrupole potential. Specifically, the optical potential is realized by the interference pattern between the two crossed arms of a bow-tie optical cavity, in order to enhance the depth of the optical trapping potential. 
If the wavelength of the laser injected into the bow-tie cavity is blue-detuned with respect to the nearest atomic transition of the ions, the optical potential will confine the ions only along the lattice direction (x axis in Figure~\ref{fig:EoScheme}), in correspondence of the minima of the interference pattern, thus minimizing the off-resonant scattering rate.
For realizing a three-dimensional trapping potential, the confinement along the two remaining axes is ensured by the static electric quadrupole, generated by two electrodes aligned along the lattice axis.
Therefore, the approximated potential in the local minimum of the optical lattice intensity pattern can be written as \cite{Perego}

\begin{equation}
\begin{aligned}
\Phi_{eo}(x,y,z)\,&=\,
\biggl( \frac{8 \, \alpha P}{\pi\, w_{0x}\,w_{0y}} \,
\biggl(\frac{1}{\tilde{w_x}^2} - \frac{4 \pi^2}{d^2} \biggr) \, + \, \frac{\kappa e_0 \phi}{R^2} \biggr) \, x^2 \, + \\
&+\, \biggl( \frac{8 \, \alpha P}{\pi\, w_{0x}\,w_{0y}} \,
\frac{1}{w_{0y}^2} \, - \, \frac{\kappa e_0 \phi}{2R^2} \biggr) \, y^2 \, + \, \biggl( \frac{8 \, \alpha P}{\pi\, w_{0x}\,w_{0y}} \,
\frac{1}{\tilde{w_z}^2} \, - \, \frac{\kappa e_0 \phi}{2R^2} \biggr) \, z^2 
\label{eq:potenziale-EO}
\end{aligned}
\end{equation}

where $\alpha$ is the (negative) state-dependent atomic polarizability, $w_{0x}$ and $w_{0y}$ are the beam radii along the x and y axes, $P$ is the power of the laser beam, $\kappa$ is a geometrical factor related to the electrodes' shape and mutual position, $R$ is the distance between the electrodes and the trap center, $\phi$ is the static electric potential applied on the electrodes, and

\begin{equation*}
 \tilde{w_x} \, = \, \biggl( \frac{\cos(\theta_c)^2}{w_{0x}^2}+\frac{\sin(\theta_c)^2}{2z_{R}^2}\biggr)^{-1/2}
 \quad \text{and} \quad
 \tilde{w_z} \, = \, \biggl( \frac{\sin(\theta_c)^2}{w_{0x}^2}+\frac{\cos(\theta_c)^2}{2z_{R}^2}\biggr)^{-1/2}
 \end{equation*}

are the rotated beam radii along the x and z directions, where $z_R$ is the Rayleigh range of the laser beam, and $\theta_C$ is the crossing angle between the diagonal arms of the bow-tie cavity.
By analogy with the anisotropic three-dimensional harmonic oscillator, the following trapping frequencies can be extracted from Equation~\ref{eq:potenziale-EO}

\begin{equation}
\begin{aligned}
\omega_{eo,\,x}^2\, &=\, \frac{16 \, \alpha P}{\pi \, m_{\text{ion}}\, w_{0x}\,w_{0y}} \,
\biggl(\frac{1}{\tilde{w_x}^2} - \frac{4 \pi^2}{d^2} \biggr) \, + \, \frac{2\kappa |e| \phi}{m_{\text{ion}}\,R^2} \\
\omega_{eo,\,y}^2\, &=\, \frac{16 \,\alpha P}{\pi\,m_{\text{ion}}\,w_{0x}\,w_{0y}} \,
\frac{1}{w_{0y}^2} \, - \, \frac{\kappa |e| \phi}{m_{\text{ion}}\,R^2} \\
\omega_{eo,\,z}^2\, &=\, \frac{16 \,\alpha P}{\pi\, m_{\text{ion}}\, w_{0x}\,w_{0y}} \,
\frac{1}{\tilde{w_z}^2} \, - \, \frac{\kappa |e| \phi}{m_{\text{ion}}\,R^2}
\label{eq:EO-frequences}
\end{aligned}
\end{equation}

\begin{figure}[t]
\centering
\includegraphics[width=0.85\textwidth]{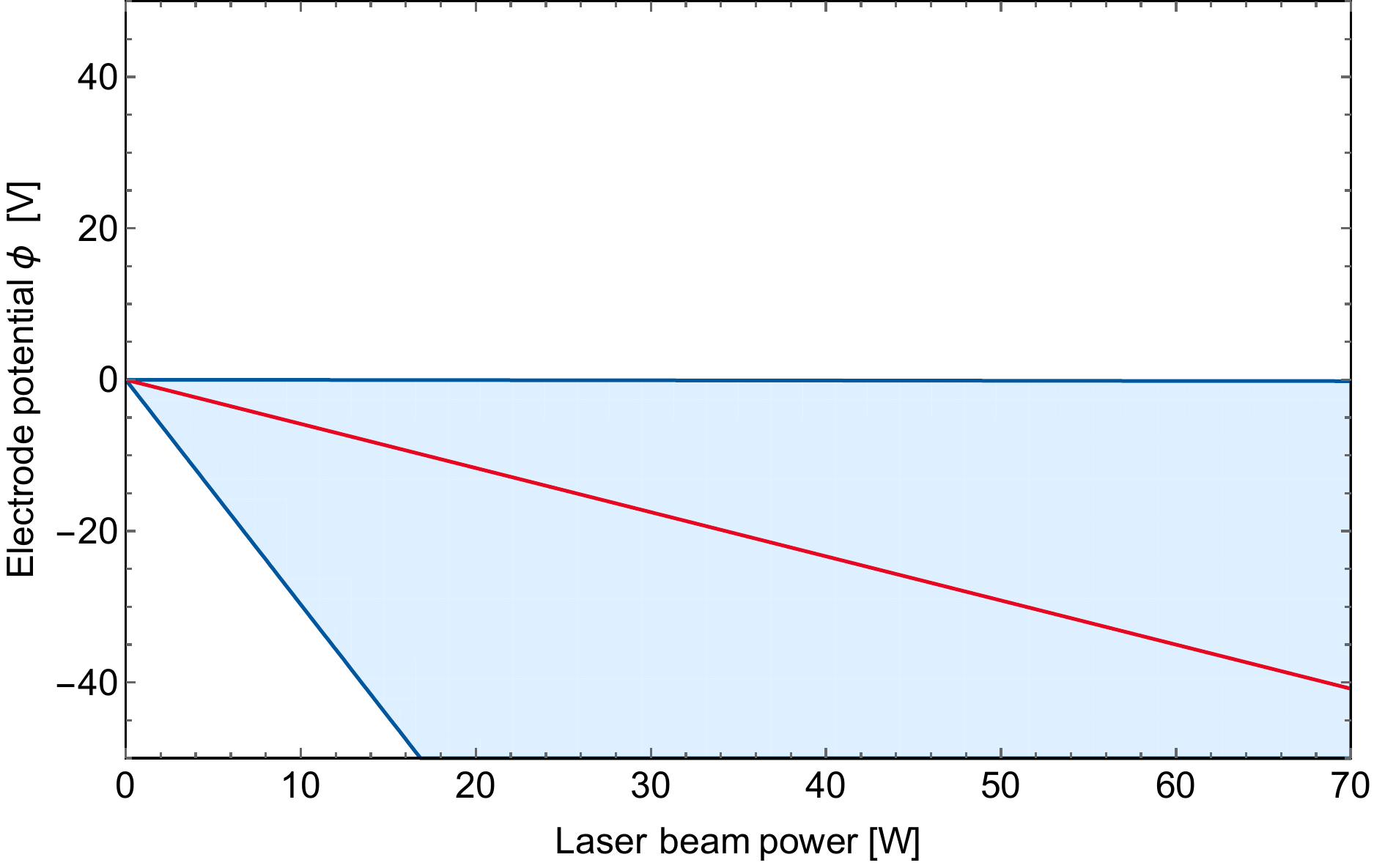}
\caption{Electro-optical trap "stability diagram". The shaded region shows all the values $\{P,\phi\}$ for which the trapping frequencies in the three directions are simultaneously positive, thus indicating a stable trapping potential. The laser wavelength, the beam waists, and the crossing angle chosen for trapping Ba$^+$ are $\lambda=451.7$\,nm, $w_{0x}=w_{0y}=40\,\mu$m, and $2\theta_c = 10\,\degree$, respectively. The red line indicates the parameters $\{P_{\text{iso}},\phi_{\text{iso}}\}$ for which the confinement is equal in each direction. The region above this line corresponds to a disk-like shape potential.
\label{fig:EOStability}}
\end{figure}

Once the trapping wavelength, the crossing angle between the two cavity arms (through which the lattice constant can be tuned), and the electrodes' geometry are fixed, the electro-optical potential depends only on the laser power $P$ and on the applied potential $\phi$. 
A "stability diagram" of the electro-optical trap can be deduced from Equations~\ref{eq:EO-frequences}: in fact, the potential acting on the ions provides a stable confinement in all three directions if the frequencies associated to the harmonic potentials are all non-negative. 
For instance, the stability plot evaluated for the trap parameters $\lambda=451.7$\,nm (a magic wavelength for the $6\,^2$S$_{1/2} \leftrightarrow 6\,^2$P$_{1/2}$ transition of Ba$^+$ \cite{sahoo}), $w_{0x}=w_{0y}=40\,\mu$m, and $2\theta_c = 10\,\degree$ is presented in Figure~\ref{fig:EOStability}.
The figure shows that the electrode potentials must be negative for ensuring an attractive potential along the y and z axes.
Moreover, the optical potential must be deep enough to prevail over the anti-trapping electric potential along the x axis. 
The red line in Figure~\ref{fig:EOStability} corresponds to the parameters $\{P_{\text{iso}}, \phi_{\text{iso}}\}$ for which there is a 3D isotropic confinement .
Therefore, for a given laser power $P$, the shape of the electro-optical potential can be tuned by changing the electric potential $\phi$ on the electrodes in the range [$0$, $\phi_{\text{trap}}$], where $\phi_{\text{trap}}$ is the largest value for which the particles are still trapped along the interference direction (x axis). 
The possible scenarios are thus
\begin{itemize}
    \item  $0 > \phi > \phi_{\text{iso}}$ The ions can be trapped in a single minima of the optical lattice, thus forming a disk-shaped crystal.
    \item $\phi = \phi_{\text{iso}}$ The three trapping frequencies are equal, so the potential is isotropic.
    \item $\phi_{\text{iso}} > \phi > \phi_{\text{trap}}$ The confinement along the interference direction is weaker than the other two. Considering the typical depth of the optical potentials and the Coulomb repulsion, the ions might lie in different minima of the optical lattice.
\end{itemize}

\subsection{Loading ions into the electro-optical trap}
In typical ion trapping experiments, ions are produced by ionizing atoms from a hot vapor created by a small oven.
Therefore, the depth of the trapping potential must be adequate ($100$s of kelvins) so that a hot ion could be confined and then laser-cooled.
In an electro-optical trap, however, the depth is limited mainly by the optical component of the potential, the depth of which is typically well below $1$\,K.
For this reason, a different strategy for loading ions in an electro-optical trap must be devised.
One possibility could be photo-ionize atoms that have already been laser-cooled in a magneto-optical trap (MOT) \cite{Sage, Cetina-Brigth}.
However, since we are planning to use Ba$^+$ ions, this solution is not practical as a Ba MOT requires three infrared repumping lights \cite{Dammalapati}.
Our plan, instead, is to use a Paul trap as a first trapping stage in which the hot ions are trapped and laser-cooled, and then transferred from the Paul trap to the EO trap. 
This strategy has the advantage of using standard ion trapping techniques for producing ions at the Doppler temperature, for which the EO trap potential is deep enough to ensure a confinement of the particles. 
In order to facilitate the swapping from the Paul trap to the EO trap, the shape of the two trapping potentials should match as much as possible. 
This request sets a number of constraints on the shape and size of the electrodes, and on the method for creating the deep optical lattice.

\section{Trap design}

Our ion trapping apparatus is composed by the electro-optical trap, the Paul trap, and the neutrals' ovens. 
These three components have to be easily integrated in the same structure, thus their design is a complex and adaptive process. In addition, another constraint on the design concerns the optical access, which must be adequate for the imaging system, the numerous laser beams for manipulating the particles and an optical transport of the neutral Li atoms towards the trap center. This necessity naturally leads to a trap with a wide trapping volume, which must also allow the crossing of the laser beams for creating the optical lattice of the electro-optical trap.

\subsection{Electro-optical trap design}

\begin{figure}[t]
\centering
\includegraphics[width=0.7\textwidth]{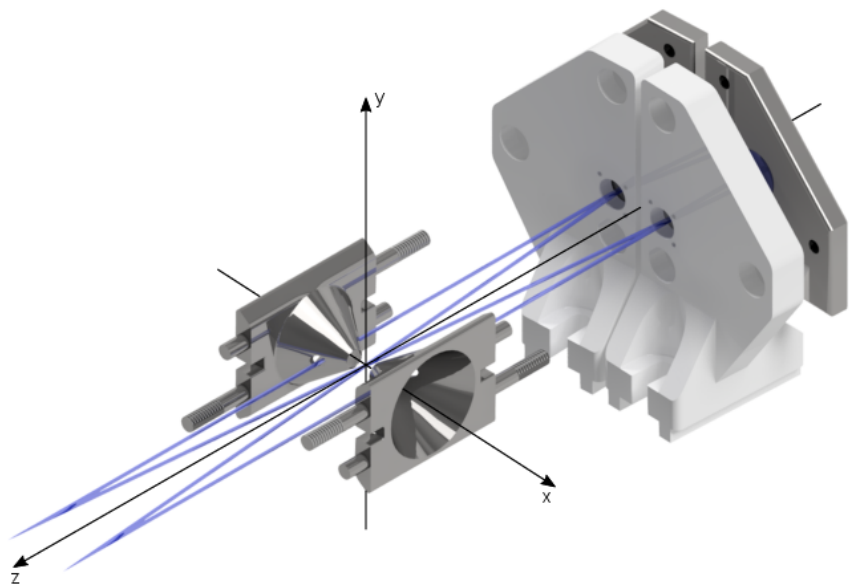}
\caption{CAD assembly of the electro-optical trap, formed by the cone electrodes and the bow-tie cavity mirror mounts. The two mirror mounts on the left side have been removed for a better view.
\label{fig:EoTrap}}
\end{figure}

Ions are trapped along two directions by a static electric quadrupole potential , whereas along the third direction (x axis) the confinement is provided by a 1D blue-detuned optical lattice. 
In our electro-optical trap, the static quadrupole potential is created along y and z axes in Figure~\ref{fig:EoTrap} by two opposite electrodes, while along the third direction (x axis) the confinement is realized by a bow-tie cavity.
Since the axis orthogonal to the 2D ion crystals (x axis) coincides with the imaging direction, the static electric quadrupole potential must be produced by a pair of ring-shaped electrodes, ensuring a circular distribution of charges with the x direction as symmetry axis.
These rings must be contained in mechanical supports that do not limit the numerical aperture of the imaging of the ion crystals' plane (y-z in Figure~\ref{fig:EoTrap}). 
To this end, we chose a conical shape for these electrodes - which we will call "cone electrodes" in this article.  
As additional constraint, shape and dimensions of the cone electrodes must be carefully adapted to the electrodes of the Paul trap.

The bow-tie cavity is based on a symmetric configuration with four curved mirrors having the same radius of curvature, so that the laser beam traveling in the cavity has a focus in each of its arms, two of which are overlapped in the center of the bow-tie. There, the two running waves create a deep lattice in which the intensity of the two interfering beams is enhanced by a factor proportional to the cavity finesse, thus increasing the trap depth \cite{devang}. 
For facilitating the process of loading the atoms and ions in a single minimum of the optical potential, the lattice constant must be as large as possible, i.e. the two lattice beams must cross at a small angle. 
In our design, the crossing angle is $10\,\degree$, the waist of the beams at the crossing is $40\,\mu$m, and the mirrors radius of curvature is $100$\,mm. 
With these parameters, the bow-tie geometry can be entirely defined in the ABCD matrix formalism by imposing the cavity stability \cite{nagourney}. The result, described in the stability diagram by the point $\{g_1, g_2\} = \{-0.89, -0.9\}$, is a strongly elongated cavity with the crossed arms $9.48$\,cm long, and parallel ones $9.45$\,cm long.

Regarding the cavity assembly, the mirror mounts are composed of two parts: a ceramic mirror holder and a metallic cover.
The mirror, which has a diameter of $6$\,mm, is attached to the metallic cover with a ultra-high vacuum compatible glue (e.g., Epoxy Technology 353ND). 
In order to tune the cavity resonant frequencies, two of the mirrors are equipped with a circular hollow piezo element (e.g. Noliac NAC2121), which is inserted between the mirror and the cover and held in position with a thin layer of glue. 
This stack is then connected to the ceramic support by using three small and finely threaded screws (M2x0.25\,mm).
An elastic element (DuPont Kalrez $8002$, an UHV compatible rubber) housed in the mirror compartment offers a contrasting force to the mirror (and the piezo, if present) which keeps the cover stable and separated from the holder even if the screws are not completely tightened. 
In this way, the mirror can be tilted by acting on the screws.
In order to facilitate the decoupling between the vertical and the horizontal degrees of freedom, the three screws are placed at the vertices of a triangle with a vertical side.
Moreover, in case one wants to rely on the mechanical precision of the assembly, a metallic hollow spacer can be inserted between the holder and the cover to fix their mutual distance, thus making the whole system ``monolithic''.

In order to realize the cavity geometry as accurately as possible, it is crucial to position the four mirror mounts on the base at the exact mutual distance and angle. To ease this task, the mirror mounts are equipped with a pair of parallel rails that can move in the complementary pair of grooves placed on a titanium base plate. 
In this way, when the mirror mounts are placed onto the base plate, they are oriented the right angle. 
Once the angular orientation is set, the position of the mounts can be set by tightening each mirror mount to the base plate with two screws. Additionally, each mirror mount and the base plate are equipped with precision holes corresponding to the exact position that the mirrors should have: by using vented dowels the mirrors can be positioned in this pre-aligned structure.

\subsection{Paul trap design}

\begin{figure}[t]
\centering
\includegraphics[width=0.85\textwidth]{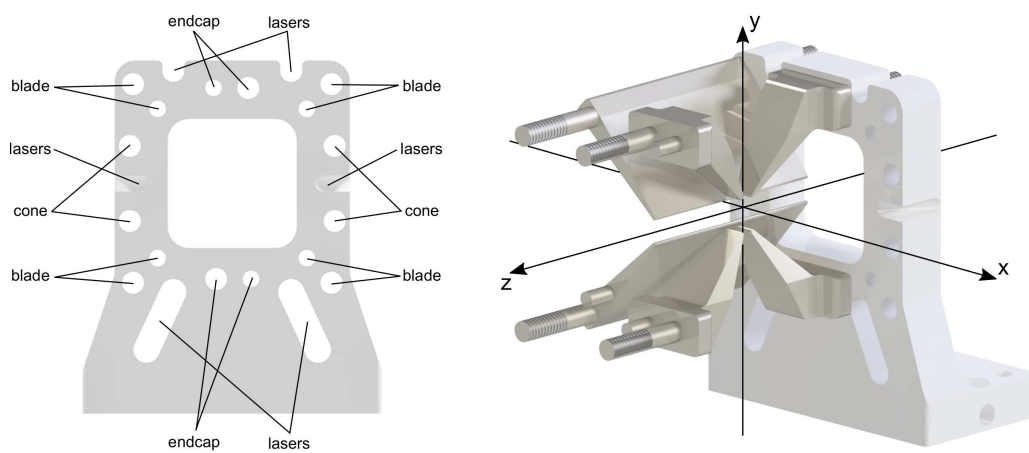}
\caption{On the left, the front view of the ceramic lateral support. Each hole is used to hold a specific electrode. Extrusions for optical access are indicated with the label "lasers". On the right, a CAD assembly of some of the electrodes (two RF blades and four endcap electrodes) inserted in the lateral support.
\label{fig:LateralSupports}}
\end{figure} 

The Paul trap is inspired to conventional linear Paul trap designs (e.g. \cite{Drewsen-trap}), with the four electrodes fed with the RF potential extended along
the z axis (refer to Figure~\ref{fig:LateralSupports}).
These electrodes are characterized by a blade shape with a very thin tip having a hyperbolic profile to better reproduce an ideal electric quadrupole. In addition, they are equipped with four pins for ensuring their precise spatial positioning.

The other electrodes of the Paul trap, the so-called “endcaps”, ensure a shallow trapping
along the symmetry axis of the RF electrodes. 
Usually, the static electric potential is realized by two electrodes places at the edges of the trapping volume \cite{Roos}. 
In this case, though, it is crucial to generate ion crystals with the shape that matches as much as possible the expected crystal shape in the EO trap, in order to optimize the transition from one trapping potential to the other. 
To this end, in order to ensure the confinement along the z axis, we designed two pairs of electrodes oriented along the y-axis (see Figure~\ref{fig:LateralSupports}). Each pair of electrode reproduces the same potential created by a hypothetical tip aligned along the y axis, but leaving a clear optical access to the ions. 
The distance between the electrodes must be small compared to the distance between them and the ions, so that particles do not experience a double-well potential. 
Otherwise, the resulting potential would take the shape of a double well, with two distinct minima along the z axis. 
With respect to their shape, these endcap electrodes are formed by a relatively large base in order to tight them to the lateral supports, while their tips are characterized by a simple circular profile.  

The electrodes of both the EO trap and Paul trap are held by a pair of ceramic supports (to which we will refer as "lateral supports"), which are placed at the sides of the trap.  
Each electrode is equipped with a pair of cylindrical extrusions (pins) in correspondence of precision holes made in the lateral supports. One in each pair of pins is partially threaded in order to tighten the electrode to the ceramic support with a nut. 
Both pins are used as dowels that facilitate the alignment and avoid any possible rotation of the electrode around its tip-screw.
In addition, the threaded tip which appears on the other side of the ceramic support is exploited to
feed the voltage to the electrode by using a tubular crimp lug tighten with a second nut.
The lateral supports are also provided with additional holes and extrusions to ensure extra optical access, as shown in Figure~\ref{fig:LateralSupports}. 
Specifically, the topmost grooves can be used to shine a laser beam towards the trap center with an angle of approximately $45\,\degree$ with respect to all the three trap axes.
After passing through the center of the trap, this laser beam passes through a second aperture in the opposite lateral support, in order to reduce stray light and avoid the laser-induced creation of charges on the ceramic surface close to the ion trap center. 
Finally, the supports have large clearance on the x-z plane which ensures a large access to the atoms.

 \subsection{The atomic source}

\begin{figure}[t]
\centering
\includegraphics[width=0.65\textwidth]{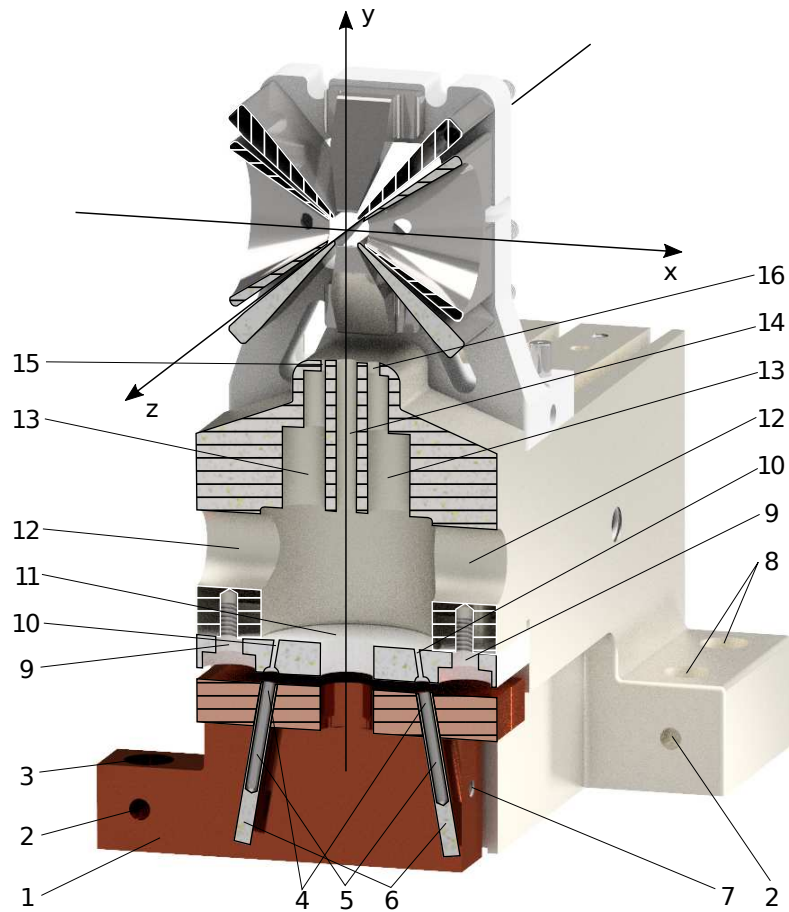}
\caption{CAD section of the ovens system for producing atomic beams of neutral Barium. (1) Copper heat-sink. (2) Venting holes. (3) Screw passing hole for fixing the heat-sink on the flange. (4) Cold upper part of the (5) tubes. (6) Tube
stripes for the electric connections. (7) Threaded hole for the ground-connecting the heat-sink. (8) Screw passing holes for fixing the base on the flange. (9) Pipes and (10) screw passing holes of the cold ceramic (11) skimmer. (12) Inspection viewports of the (13) cylindrical oven cavities. (14) Anti-sputter wall for protecting the vertical optical access. (15) and (16) Output holes.
\label{fig:OvensSection}}
\end{figure}

Two independent ovens for producing vapors of neutral atoms are integrated into the metallic base, below the trap electrodes. 
The overall structure is formed by two tubes (element 5 in  Figure~\ref{fig:OvensSection}), a copper heat-sink~(1), a cold ceramic skimmer~(11), and the metallic base itself.
The tubes are tubular pipes in which the solid metallic element is crammed.
By applying to the thin stripe~(6) at their bottom a high current for a short time interval, the heat produced via Joule dissipation warms the ovens up until a stream of atomic vapor is emitted. 
Therefore, the tubes design is based on the necessity to facilitate the dissipation, thus the produced heat. 
For this reason, the ovens are produced in stainless steel, a metal characterized by a low thermal conductivity as compared to other metals. 
In addition, the electrical resistance is increased by minimizing the tubes' thickness (internal and external diameters are respectively $1.20$\,mm and $1.50$\,mm).
The tubes are fixed into a copper heat-sink, which has two roles. First, since the heat-sink can be considered in thermal equilibrium with the
cold flange to which it is fixed with the screws~(8), it rapidly lowers the temperature of the tubes after the production of the atomic shot, starting from the cold upper side~(4). Second, the heat-sink is connected to grounded wires, hence offering to the current a path to flow.
The atomic beams, which can be independently activated, are oriented towards the trap center. 
During their flight, the neutral atoms pass through a trimming system composed by the skimmer and the metallic base.
The skimmer has a pair of tilted pipes~(10) - about $2.5$\,mm long, with a diameter of $0.5$\,mm - which reduce the solid angle of the atomic beam. 
The metallic base, to which the skimmer is attached, is provided with an inner cavity dug via a series
of tangent cylindrical removals~(13), which culminate with two output holes having different diameters, $0.4$\,mm~(15) and $1.2$\,mm~(16). 
In particular, the smaller oven is calibrated to reach the trap center and avoid detrimental electrodes sputtering.
Additionally, the optical vertical access to the trap is protected by a cylindrical wall~(14), thus avoiding that the atomic jets could pass thorough it. 
The mutual alignment of the ceramic skimmer and the base plate is a fundamental requirement in order to ensure the creation of a velocity-selected atomic beam passing by the center of the trap. 
To this end, the correct positioning of the skimmer is ensured by two dowels.

\subsection{Machining tolerances}
The successful assembly of a composite trap relies on achieving high machining precision. 
In our assembly, the most delicate points in which a high fabricating precision is strongly required are the rail-groove system providing the orientation of the mirror holders, the blade and cone electrodes, and the external diameter of the oven tubes, which must be inserted in the heat-sink and hold by friction. 
Additionally, dowels are particularly critical, since they represent the means extensively used to ensure the precise positioning of the most critical parts.
On the most demanding dimensions, a tolerance of $\pm 5\,\mu$m or better was reached. 
After the production, each single trap component was individually measured by using an optical microscope and a coordinate measuring machine (CMM), a device able to measure the size of an object with a $\mu$m-level of precision.  
This measurements procedure ensures that all the produced parts are optimal reproductions of the ideal objects described by the technical drawings within the specified tolerances. 
Nevertheless, even if all the trap components were machined within the tolerances, other detrimental deviations from the ideal assembly may arise due to the tolerance interplay when different items are assembled together. 
For instance, an alignment dowel can be produced with a diameter slightly bigger than its ideal value, but still being within the tolerances specified to the constructor, and the corresponding hole into which it should be inserted can be within tolerances as well, but slightly smaller. If this is the case, the dowel may not enter into the hole, making both parts useless. 
In order to avoid these unfortunate situations, all the critical points of the whole trap assembly - mainly the dowels inserting, the pairing between the mirror mount rails and the base grooves, the electrodes alignment, and the tubes joint with their heat-sink - must be treated with special care during the construction process, and the part features should be adaptively machined, if possible. 
Moreover, more copies of these parts were realized and characterized with the CMM, so that only the best combination of matching items is selected for the trap assembly.
After assembling the whole trap, we measured the position of the electrodes with the CMM and found that all the discrepancies between the CAD design and the actual realization were $<\,10\,\mu$m, with one outlier at $14\,\mu$m.

\subsection{Materials}
In the ion trapping community there is not a definitive choice for the best material for producing the electrodes, and different options like beryllium copper, molybdenum, gold-plated alumina, etc. - have successfully been used \cite{Peik-materials}.
For our electrodes, we opted for titanium (alloy Ti6Al4V, also known as "titanium grade $5$"), mainly because it can be easily machined and a lower thermal expansion coefficient compared to other metals.
Other metallic parts are produced in titanium, as dowels and the trap base plate. Regarding the insulating material, we chose an aluminum nitride ceramic commercially known as "Shapal Hi-M Soft". 
This ceramic has the great advantage of combining a high thermal conductivity ($92$\,W/mK) with a low thermal expansion coefficient (about $4.8\,\mu$m/mK).
The trap lateral supports and the cavity mirror holders are machined in Shapal, since this ceramic can ensure a good dissipation of the heat that can be generated by the trap. 

\section{Simulations on the trapping system}


\subsection{Electrical simulations}

The design of the trap electrodes is a result of a series of numerical simulations of the electric potential, used to optimize the trap components step-by-step.
The simulations are based on the static problem approach, meaning that the varying electric fields are assumed to slowly change with respect to the electromagnetic wavelength. 
Within this approximation, the problem reduces to solve the Laplace equation, which has been numerically treated in a script written in C language using the toolkit ROOT developed by Cern \cite{root} and the package BEMSOLVER \cite{Singer}. 
The latter is a library in which the algorithms required to solve electrostatic problems with complex charges distributions are already implemented and optimized.
The approach of BEMSOLVER is based on the boundary element method (BEM), which converts a volume problem into a surface one. 
The basic idea underneath this method is to consider only the surfaces of the objects, then to find a continuous electric charge distribution on these surfaces that satisfies the boundary conditions of the original problem.
The numerical solution of the superficial charge density can be found by cutting the interfaces between different media in squared or triangular panels (mesh), small enough to consider uniform the charge density on the surfaces. 
At this point, the electric potential can be expressed as

\begin{equation}
 \phi(\vec{r}\,) \, = \, \frac{1}{4 \pi \varepsilon_0} \, \sum_{i=0}^N\,\biggl( \int_{S_i} \frac{1}{|\vec{r}-\vec{r}\,'|}\, dS'\,\biggr)\, \sigma_i
\label{eq:inverting-potential}
\end{equation}

where $\sigma_i$ are the i-th superficial charge densities of the surfaces $S_i$ and $N$ is the panels number.
Equation~\ref{eq:inverting-potential} can be reformulated in a compact way as $A \cdot \vec{q} = \vec{p}$, where the $N \times N$ matrix $A$, which depends only on the mesh geometry, connects the vector $\vec{q}$ containing the superficial charge distributions $\sigma_i$ to the vector $\vec{p}$ representing the applied voltage on the i-th panel.
BEMSOLVER employs two techniques for solving Equation~\ref{eq:inverting-potential}. 
The first is the generalized minimum residual (GMRES) method \cite{saad}, which aims at computing an approximate solution by running an iterative algorithm until a certain tolerance fixed by the user is reached. 
This method starts with an initial guess $\vec{q}_0$ for the solution, then evaluates the first residual $\vec{r}_0 = \vec{p} - A \vec{q}_0$, on which the convergence
of the algorithm is checked. 
If another iteration is needed, the new solution vector is computed as

\begin{equation*}
 \vec{q}_{i+1} \, = \, \sum_{j=0}^{i} \alpha_j\vec{q}_j + \beta \vec{r}_i
\end{equation*}

to minimize the new residual $\vec{r}_{i+1}$ by properly choosing the coefficients $\alpha_j$ and~$\beta$.
The second employed method is the fast multipole method (FMM) \cite{Greengard}, which basically describes the potential in a certain point with the single contribution due to the nearest charge, whereas the effect of the remaining charges is evaluated as a perturbation given by multipole and local expansions.

\subsubsection{Paul trap stability diagram}

\begin{figure}[t]
   \centering
   \includegraphics[width=0.80\textwidth]{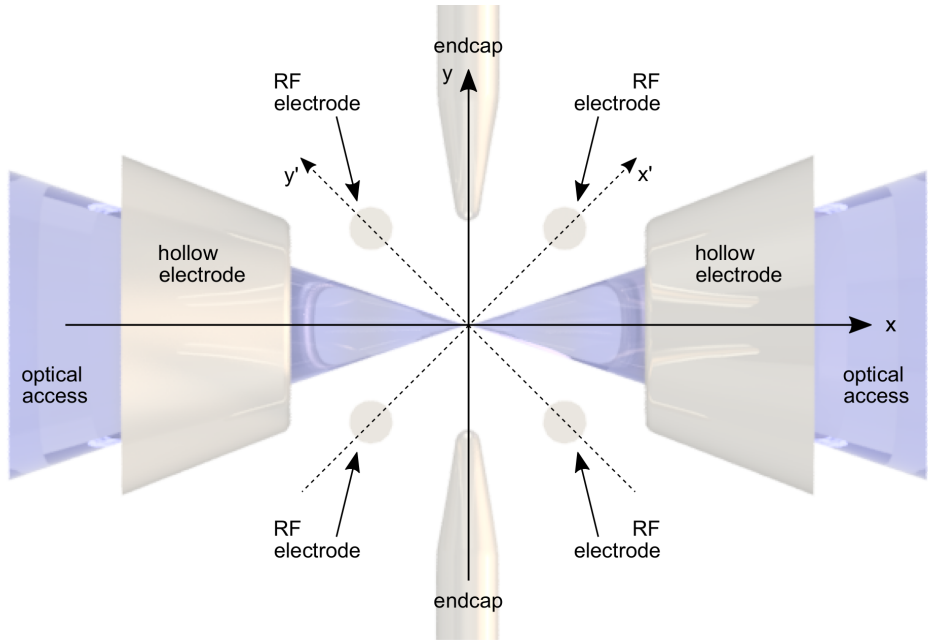}
   \caption{Overview of the Paul trap electrodes position in relation with the electro-optical trap cone electrodes. The RF electrodes are rotated of $45\degree$ with respect to the endcaps' frame, and the confinement along the z axis is provided by the negatively charged endcaps.}
   \label{fig:traps-toymodel-geometry}
\end{figure}

We used the simulation to extract the geometric coefficients linking the electric field in the center trapping region to the voltages applies to the electrodes. 
Once the trap geometry has been characterized, the electrical behavior of the electrodes can be predicted, and the stability diagram of the Paul trap can be evaluated.
It is important to note that the "blade" electrodes are rotated with respect to the endcap electrodes’ axes by an angle of $\theta = 45\degree$ (see Figure~\ref{fig:traps-toymodel-geometry}).
This rotation implies that the operation of the Paul trap is described by a system of two coupled Mathieu equations \cite{Drewsen-stability}.
The expression of the dynamic electric potential, with the coordinates rotated from $\{x',y'\}$ to the $\{x,y\}$ frame, can be written as

\begin{equation*}
 V_{RF}(x,y) \, = \, -\frac{e_0 \phi_{RF}}2 \frac{\kappa_{RF}}{R_{RF}^2}\, \bigl( (\cos(\theta)x + \sin(\theta)y)^2 - (\cos(\theta)y - \sin(\theta)x)^2\bigr) \cos(\omega_{RF} \, t)
\end{equation*}

where $\theta$ is the coupling angle, $\phi_{RF}$ is the amplitude on the potential applied on the blades, $\omega_{RF}$ is the RF pulsation, $r_{RF}$ is the distance of the blade electrode from the center of the trap, and $\kappa_{RF}$ is a geometrical factor smaller than unity that takes into account the electrodes’ shapes and shielding effects.
Taking into account the contributions of both the dynamical and static electric potentials and the definition of the dimensionless Mathieu parameters $\vec{q}$ and $\vec{a}$

\begin{equation*}
\begin{aligned}
 \vec{q} \,&= \, q
 \begin{pmatrix}1, & -1, & 0 \end{pmatrix}
 \, = \,
 \frac{2e_0\,\phi_{RF}}{m_{\text{ion}}\, \omega_{RF}^2}\frac{\kappa_{RF}}{R_{RF}^2}
 \begin{pmatrix}1, & -1, & 0 \end{pmatrix}
\\
 \vec{a} \, &= \, a \, 
 \begin{pmatrix}-1, & -1, & 2 \end{pmatrix}
 \, = \,
 \frac{4e_0\,\phi_{DC}}{m_{\text{ion}}\, \omega_{RF}^2}\frac{\kappa_{DC}}{R_{DC}^2}
 \begin{pmatrix}-1, & -1, & 2 \end{pmatrix}
\end{aligned}
\end{equation*}

the coupled motion equations for the x and y axes are

\begin{equation}
\begin{cases}
\;\ddot{x} \, = \, 2 \, q \, \bigl(\cos(\theta)x+\sin(\theta)y \bigr) \, \cos(2\tau) - a \, x \\
\;\ddot{y} \, = \, 2 \, q \, \bigl(\sin(\theta)x-\cos(\theta)y \bigr) \, \cos(2\tau) - \alpha  \, a \, y
\label{eq:system}
\end{cases}
\end{equation}

where $\alpha$ is a geometrical factor expressing the ratio  between $a_x$ and $a_y$, i.e. $a_x = a_y/\alpha = a$. The substitution $\omega_{RF} t = 2\tau$ was also performed in order to make the equations time-dimensionless. 
The stability condition of this system can be evaluated by applying the Floquet theorem. To do so, Equations~\ref{eq:system} must be first expressed as a system of first order differential equations, i.e. in matrix notation  $\dot{\vec{x}} \, = \, B(\tau) \, \vec{x}$, where $\vec{x} = (x, y, v_x, v_y)^T$ and the periodic matrix $B(\tau) = B(\tau + n\pi)$ ($n$ integer number) is 



\begin{equation*}
B(\tau) \, = \, 
\begin{pmatrix}
0& 0 & 1& 0 \\ 
0& 0 & 0& 1 \\
2q\cos(\theta)\cos(2\tau) - a & 2q\sin(\theta)\cos(2\tau) & 0& 0 \\
2q\sin(\theta)\cos(2\tau) & -2q\cos(\theta)\cos(2\tau) - \alpha a & 0& 0 \\
\end{pmatrix}
\end{equation*}

The Floquet theorem states that the differential equations' system has at least one solution of the form~\cite{Likins}

\begin{equation}
 x^{(j)} (\tau)\, = \, \exp\biggl(\,\log(\lambda_j)\,\frac{\tau}{\pi}\biggr) \, p_j(\tau) \quad \text{with} \quad p_j(\tau + \pi) = p_j(\tau)
 \label{eq:Floquet-solutions}
\end{equation}

where $\lambda_j$ are the eigenvalues of the matrix $Z(\pi)$ defined as

\begin{equation}
 \dot{Z}(\tau) =  B(\tau) \,Z(\tau) \quad \text{with} \quad Z(0) =   \mathbb{I}_{4}
\label{eq:numerical-equation}
\end{equation}

The solution Equation~\ref{eq:Floquet-solutions} can also be reformulated as 

\begin{equation*}
  x^{(j)} (\tau)\, = \, \exp\biggl(\,\log(|\lambda_j|)\,\frac{\tau}{\pi}\biggr) \, \exp(i\arg(\lambda_j))\, p_j(\tau)
\end{equation*}

from which is evident that the absolute value of the eigenvalues $|\lambda|$ must be less than unity to obtain a stable solution. 
Therefore, in order to ascertain if the trap is stable for a pair of parameters $\{q,a\}$, one has to numerically solve Equation~\ref{eq:numerical-equation} and check the absolute value of the $Z(\pi)$ eigenvalues.
The disadvantage of this procedure is that the stability boundaries are not well defined. 
To overcome this inconvenient and find an approximated expression for the stability boundaries, one can use the multiple-scale perturbation theory~\cite{Ozakin}. 
Basically, this approach consists in substituting an independent variable with a pair of fast-scale and slow-scale independent variables. 
Then, the additional degree of freedom is exploited to remove long-term and non-periodic variations in the approximated solutions, which usually limit their validity only in short time range when standard perturbation techniques are used instead.  
For a system of coupled Mathieu equations, the approximated boundaries of the first stability region are

\begin{equation*}
 \begin{aligned}
  a_1\, &= \, -\frac{1}2 q^2 \\
  a_2\, &= \, \frac{1}{2\alpha} q^2 \\
  a_3\, &= \, 1-\cos(\theta)q -\biggl( \frac{\cos(\theta)^2}8 + \frac{2\sin(\theta)^2(5+\alpha)}{(1+\alpha)(9+\alpha)} \biggr)q^2 \\
  a_4\, &= \, -\frac{1}\alpha \biggl(1-\cos(\theta)q -\biggl( \frac{\cos(\theta)^2}8 + \frac{2\sin(\theta)^2(5+1/\alpha)}{(1+1/\alpha)(9+1/\alpha)} \biggr)\,q^2\biggr)
 \end{aligned}
\end{equation*}

\begin{figure}[t]
\centering
\includegraphics[width=0.85\textwidth]{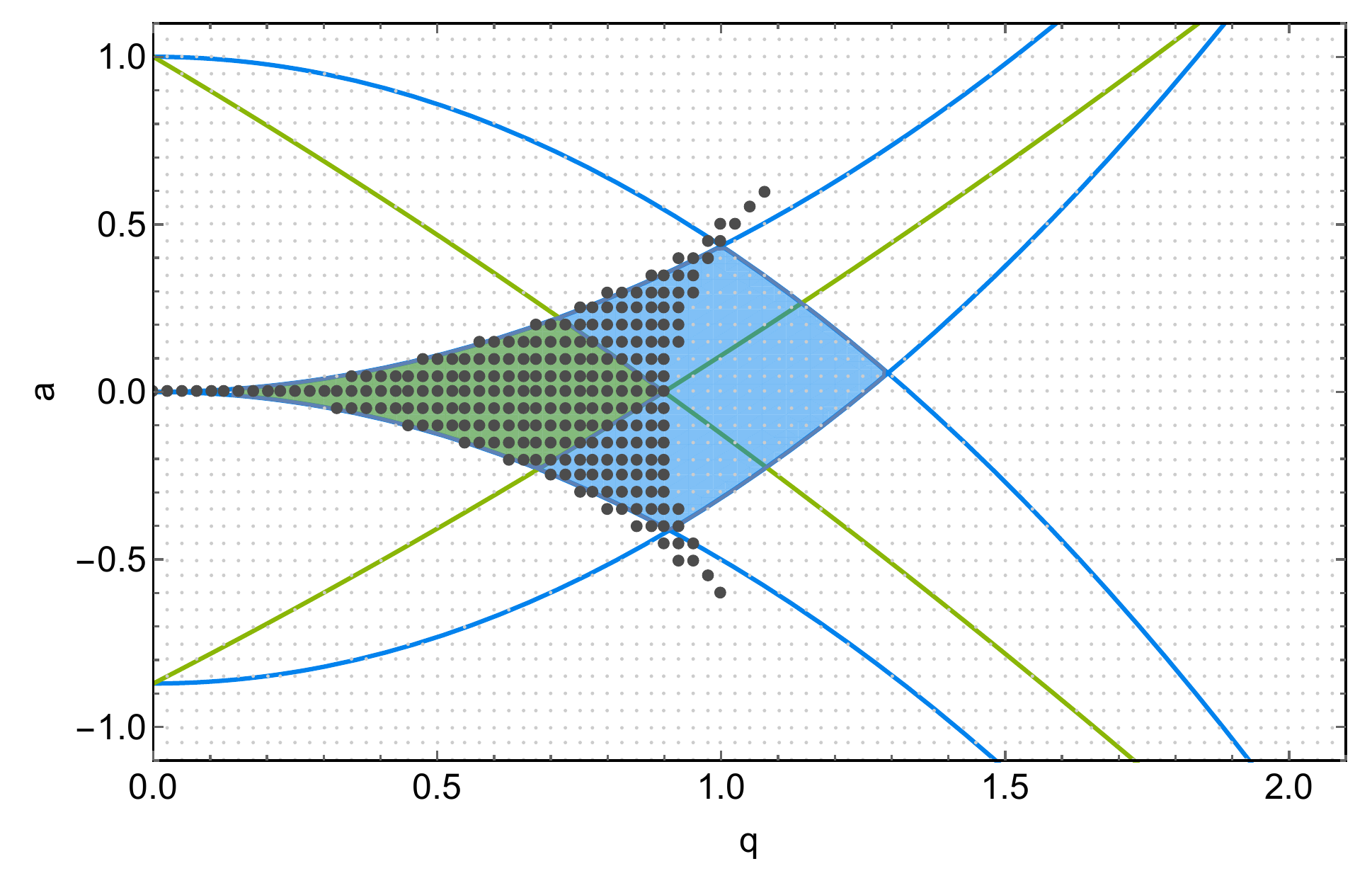}
\caption{Stability diagram of the Paul trap, characterized by a coupling angle of $45\,\degree$ between the RF electrodes and endcaps axes. 
The black thick dots represent the $\{q, a\}$ pairs for which the Paul trap (described by a system of coupled Mathieu equations) is stable, whereas the small gray dots indicate values of parameters for which the trap is unstable. 
The blue curves, calculated with the multiple-scale perturbation theory, enclose a stability region which is wider than the numerically estimated one, yet well describing the stability behavior for small values of q. The
green area is the stability region of the corresponding uncoupled system.
\label{fig:PaulStability}}
\end{figure}

The stability diagram of the Paul trap evaluated by using both numerical solutions and multiple-scale perturbation theory is reported in Figure~\ref{fig:PaulStability}.
The green area represents the first stability region for the uncoupled Mathieu equations ($\theta = 0$) and the geometry coefficients of our trap.
The first stability region evaluated with the numerical method (black thick dots) is well approximated by the boundaries obtained
with the multiple-scale perturbation theory (blue curves) only for small values of $q$. 
For $q > 0.7$, the boundary curves enclose a region of $\{q, a\}$ pairs (depicted in blue) for which the numerical method predicts instability. 
For small $q$, both coupled and uncoupled systems have stability regions characterized by the same boundaries; nevertheless, the coupled system has in general a wider stability region. 
However, while the primary stability region does not appreciably vary for small $q$ with nonzero mixing angles $\theta$, the presence of a coupling term strongly affects the secondary stability region. 
When $\theta = 45\degree$, as in our trap, the maximum coupling is reached: the result is that the secondary stability regions shift towards the primary one and join it forming an exceptional wider triangular-shaped stability region \cite{Ozakin}, as shown by the black dots in the graph of Figure~\ref{fig:PaulStability}.
There is a discrepancy between the numerical and the analytic calculations. Nevertheless, Paul traps are usually operated in the lower part of the stability diagram, where both methods predict a stable operation of the trap. For the coefficients of our trap geometry, a potential difference on the blade electrodes of $200$\,V with $\omega_{RF} = 3.2\,$MHz corresponds to $q \approx 0.15$ \cite{Detti}.

\subsubsection{Residual axial radiofrequency}
In the simulations, the trap electrodes have been considered as perfectly machined and aligned. Nevertheless, this is an ideal case, since misalignments or imperfections may occur during the fabrication or the assembly. 
Regarding the Paul trap, for instance, any deviation of the blade electrodes from the ideal case causes the presence of a nonzero RF field along the axial direction.
The approximate solution to the equation of motion for a particle with mass $m$ and charge $+e_0$ in a Paul trap in presence of an additional DC field can be written as \cite{Berkeland}  

\[
 x_i(t) \, \approx \, 
 \biggl( x_{0, i}+x_{1, i}\cos(\omega_i t + \varphi_i) \biggr)  \, 
 \biggl( 1 + \frac{q_i}2 \cos(\omega_{RF}t)\biggr) \quad \text{with} \quad i=x,y,z
\]

where $x_{1,i}$ is the equilibrium position of the ion along the i-th direction, $x_{0,i}$ is the position shift from $x_{1,i}$ due to stray electric field, $\omega_{RF}$ is the RF frequency, and $\omega_i $ is the secular motion frequency. 
Consequently, the micromotion amplitude along the i-th direction can be estimated as $(x_{0,i} \, q_i)/2$.
Ideally, along the RF electrodes' axis, the parameter $q_z$ is null near the trap center; instead, it assumes nonzero values if finite electrodes length and misalignments effects are introduced. 
In order to evaluate $q_z$ and the residual axial micromotion amplitude, a modified version of the trap assembly, having the blade electrodes displaced from their exact position, was simulated.
Since the mechanical tolerance on the blades' position is $5\,\mu$m, the angular deviations were neglected in the simulation, while the linear displacements were considered equal to the tolerance for taking into account the worst possible case. 
This evaluation shows a residual axial micromotion amplitude of about $0.5$\,nm, which corresponds to a micromotion energy of about $1\,\mu$K.

\subsection{Thermal simulations}


\subsubsection{RF power dissipation}
Each pair of electrodes can be schematized as a capacitor having a complicated shape and containing a dielectric medium (ceramics of the lateral supports) disposed only around their cylindrical extrusions.
As a result, an equivalent series resistance (ESR) can be associated to each electrode, the value of which can be estimated as

\[
ESR \, = \, \frac{\tan(\delta)}{\omega_{RF} \, C}
\]

where C is the lossless electrode capacitance and $\tan(\delta)$ is the loss tangent, which arises from the phasor representation of the equivalent circuit parameters. For Shapal Hi-M Soft, its value is ${\tan(\delta) \approx 10^{-3}}$.
The reactive term of the $ESR$ is responsible for the heating of the parts composing the trap by Joule effect. 
The electrodes' $ESR$ is calculated by using a COMSOL Multiphysics simulation to estimate the corresponding capacitances $C$.
In the case of the blade electrodes, with ${\omega_{RF} = 2 \pi \cdot 3.5\,\text{MHz}}$ and an RMS amplitude of $200$\,V/$\sqrt{2}$, we found an overall dissipated power of $0.69$\,mW for the upper blades and $0.74$\,mW for the lower ones. 
These values can be employed for simulating the trap heating due to the RF power losses in stationary conditions, using as the only constraint a fixed temperature for the external side of the flange ($293$\,K).
The results showed a small temperature increase on the order of tens of mK, with the hot spots mainly localized at the interfaces between the electrodes and the lateral supports. Regarding the thermal perturbation on the optical cavity, the simulation shows at most a vertical thermal expansion of $0.4$\,nm for the ceramic lateral supports, while along the directions in the horizontal plane there are no appreciable expansions.

\subsubsection{Ions loading from Paul trap to electro-optical trap}

Studying the temperature variation of the trap is fundamental for estimating possible changes of bow-tie cavity length. In particular, in the transition phase from the Paul trap to the electro-optical trap, a local heating source (dissipated RF power) is turned off, thus possibly causing instabilities and long thermal drifts to the bow-tie cavity and the trapping interference pattern.
If the temperature of the system undergoes a decrease of about $2$\,K, the calculations show that the corresponding variation of the optical cavity length would lead to a shift of the central minimum of the interference pattern equal to the waist of the trapping laser beam ($40\,\mu$m). Therefore, the trapped ions would be slowly displaced until reaching the end of the trapping region.
For better stabilizing the system temperature and avoiding shifts of the confining interference pattern, a plate made in copper is attached to the titanium base plate of the trap.
This element is thermally connected to outside the vacuum chamber through a thick feedthrough fixed on one of its extremities. 
In this way, the plate works as a "cold finger" that helps dissipating the heat produced in the trap and, possibly, actively control the trap temperature. To this end, a vacuum-compatible thermistor is placed on the plate to monitor the trap assembly temperature.

\subsubsection{Ovens' simulations}

\begin{figure}[t]
\centering
\includegraphics[width=0.85\textwidth]{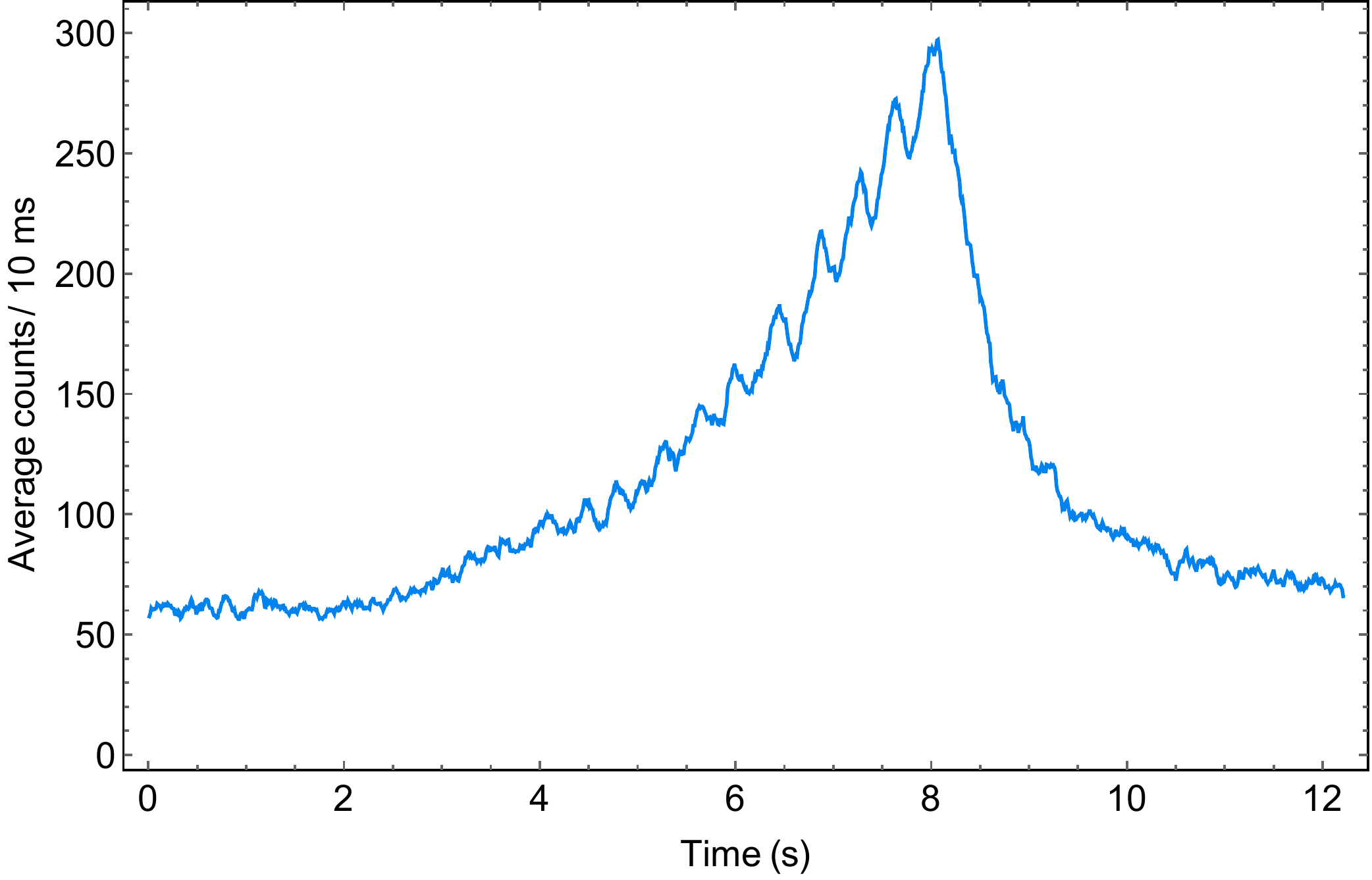}
\caption{Average counts of photons emitted by neutral Barium atoms decaying from 5d6p $^3$D$_{1}$ level are plotted as a function of time. The neutrals are emitted from one of the two ovens by applying a train of $20$ current pulses $400$\,ms long with a duty cycle of $75$\%, whereas the amplitude is linearly decreased from $100$\,A to $70$\,A to keep the temperature from exponentially rising. Time $t=0$ corresponds to the moment in which the current pulses' train starts.
\label{fig:BaFromOvens}}
\end{figure}

Thermal simulations with COMSOL Multiphysics on the ovens' system are essential for establishing the behavior of the tubes when high current bursts are applied.
The parameters of the current pulses' train are fundamental knobs that have to be experimentally adjusted in order to reach a satisfying trade-off between the number of the atoms in the vapor flux and the minimization of the vacuum pollution.
For instance, the response of the ovens to a given current can be characterized by detecting the neutrals' flux.
The reported trapping system has been designed for an experiment that uses Barium ions, generated through a two-photons resonant photo-ionization process \cite{Leschhorn_2012}. 
Neutral Barium atoms coming out of the ovens are firstly excited through the intercombination line from the ground state 6s$^2$ $^1$S$_{0}$ to the level 5d6p $^3$D$_{1}$ by shining light at $413.2$\,nm, then they are ionized by absorbing a second photon at the same wavelength. 
However, Barium atoms in the intermediate energy level can decay to the 6s5d $^3$D$_{1}$ or 6s5d $^3$D$_{2}$ metastable levels emitting a photon at $659.5$\,nm or $667.5$\,nm, respectively. 
Therefore, it is possible to detect the atomic flux out of the ovens by exciting the atoms with blue light and detecting their fluorescence in the red region of the spectrum.
The graph displayed in Figure~\ref{fig:BaFromOvens} reports the average photon counts for a binning time of $10$\,ms, feeding one of the two tubes with a train of $20$ current pulses $400$\,ms long with a duty cycle of $75$\%, linearly modulated in amplitude with a decreasing trend from $100$\,A to $70$\,A. The measurement was taken by using the photon counting head Hamamatsu H11870-02 and the photon counting unit Hamamatsu C8855-01. 
The graph, where $t=0$ marks the beginning of the current feeding, shows that the first current pulses just heat up the tube without causing the appreciable neutrals' emission, and that the atomic flux begins in the second half of the pulses' train. When the current is switched off (after $8$\,s), the emission rapidly decreases due to the copper heat-sink draining out the heat. 
At the end of the sequence, the neutrals' emission spikes caused by the last pulses of the sequence can be recognized. 

Another aspect that plays a central role in the optimal functioning of the ovens is the geometry, and in particular the ratio between the internal and external diameters $\phi_{\text{int}}/\phi_{\text{ext}}$. 
Increasing this ratio, i.e. reducing the tube section, causes an increment of the dissipated heat via Joule effect. Nevertheless, the diameters' ratio is limited by the feasibility of its mechanical fabrication (in the presented case, this ratio is $0.8$). 
With lower values, thermal simulations show a temperature increasing during the current burst much less than the one predicted for ${\phi_{\text{int}}/\phi_{\text{ext}}=0.8}$, thus leading to longer current pulses in order to obtain proper atomic vapors.

\section{Conclusions}

In this work, we have reported the design of an electro-optical trap for ions, suitable for experiments with atom-ion quantum mixtures.
The trap assembly, composed by a linear Paul trap, an electro-optical trap and two ovens, results from an adaptive design process strongly based on numerical simulations for determining the shape of the electrical potentials and the heat distribution on the electrodes.
This trap has specifically been designed for a quantum mixture of Barium ions and Lithium atoms. Nevertheless, the apparatus can be easily adapted to other atom-ion mixtures with minor changes of the optical setup.
The position and shape of the Paul trap's endcaps ensure a wide optical access to the ion trapping region, facilitating the optical transport of ultracold atoms, which will be produced in an optical trap in a separate vacuum chamber \cite{berto2019prospects}. 
The production of quantum gases and trapped ions will be synchronized by managing the overall experimental sequence with a single control apparatus \cite{Perego-paper}.
The micromotion-free dynamics of the ions in the electro-optical trap will allow the ions to reach ultracold temperatures by sympathetic cooling with a quantum gas, eventually leading to atom-ion collisions in the s-wave scattering regime.


\vspace{6pt} 




\funding{This work was funded by the ERC Starting Grant PlusOne (Grant Agreement No. 639242), the project EMPIR 17FUN07 (CC4C), the SIR-MIUR grant ULTRACOLDPLUS (Grant No. RBSI14GNS2), and the FARE-MIUR grant UltraCrystals (Grant No. R165JHRWR3). 
This project has received funding from the EMPIR programme co-financed by the Participating States and from the European Union’s Horizon 2020 research and innovation programme.}

\acknowledgments{
We thank A.Detti and F. Berto for experimental assistance and M. Inguscio for continuous support. }

\conflictsofinterest{The authors declare no conflict of interest.}
\reftitle{References}


\externalbibliography{yes}
\bibliography{Bibliography.bib}



\end{document}